\documentclass[12pt,preprint]{emulateapj}

\newcommand{\be}{\begin{equation}}
\newcommand{\ee}{\end{equation}}
\newcommand{\ba}{\begin{eqnarray}}
\newcommand{\ea}{\end{eqnarray}}

\def\Mpc{{\rm Mpc}}

\def\GHz{{\rm GHz}}

\def\cm{{\rm cm}}

\begin{document}

\title{The OH line contamination of 21 cm intensity fluctuation measurements for $z=1\sim4$}

\author{Yan Gong$^1$}
\author{Xuelei Chen$^{2,3}$}
\author{Marta Silva$^4$}
\author{Asantha Cooray$^1$}
\author{Mario G. Santos$^4$}
\affil{$^1$Department of Physics and Astronomy, University California, Irvine, CA92097, USA}
\affil{$^2$National Astronomical Observatories, Chinese Academy of Sciences, Beijing, 100012, China}
\affil{$^3$Center of High Energy Physics, Peking University, Beijing, 100871, China}
\affil{$^4$CENTRA, Instituto Superior Tecnico, Technical University of Lisbon, Lisboa 1049-001, Portugal}

\begin{abstract}
The large-scale structure of the Universe can be mapped with unresolved intensity fluctuations of the 21 cm line.
The power spectrum of the intensity fluctuations has been proposed as a probe of the  baryon acoustic oscillations at 
low to moderate redshifts with interferometric experiments now under consideration.
We discuss the contamination to the low-redshift 21 cm intensity power spectrum generated by the 18 cm OH line 
since the intensity fluctuations of the OH line generated at a slightly higher redshift contribute to the intensity fluctuations observed in an experiment. We assume the OH megamaser luminosity is correlated with the star formation rate, 
and use the simulation to estimate the OH signal and the spatial anisotropies. We also use a semi-analytic simulation to
predict the 21 cm power spectrum. At $z=1$ to 3, we find that the OH contamination could reach 0.1 to 1\% of the 21 cm
rms fluctuations at the scale of the first peak of the baryon acoustic oscillation.
When $z>3$ the OH signal declines quickly, so that the contamination on the 21-cm becomes negligible at high redshifts.
\end{abstract}


\maketitle

\section{Introduction}

The large-scale structure of the Universe can be observed efficiently with the 
{\it intensity mapping} technique, where the distribution of the 
radiation intensity of a particular line emission 
from large volume cells is observed without attempting to resolve
the individual emitters, galaxies, within the volume. This technique is
particularly suitable for radio observations, where the angular resolution 
is relatively low. By observing the radiation at different wavelengths,  the emissivity from different 
redshifts are obtained, thus revealing the three dimensional
matter distribution on large scales. It was first recognized that 
this method can be applied to the 21cm line of the 
neutral hydrogen \citep{Chang08,Peterson09,Chang10}, and provides 
a very powerful tool for precise determination of the equation of state of 
dark energy by using the baryon acoustic oscillation (BAO) peak of large-scale structure 
as a standard ruler \citep{Chang08,Ansari08,Seo10}. 
More recently, it has also been proposed that the intensity mapping method be used for
molecular and fine-structure lines, such as
CO \citep{Gong11,Carilli11,Lidz11,Visbal10} and CII \citep{Gong11b}.

A possible problem with the intensity mapping technique is the contamination 
by other lines. Unlike the observation of individual sources, where different lines from two different sources overlapping
along the line of sight can be separated through high-resolution imaging,
in intensity mapping the contamination from a different line at a  different redshift cannot be easily separated. 
For optical (e.g. the Lyman $\alpha$ line) and for many of the important radio lines,
there are many spectral lines with wavelengths longer than the line being observed, 
thus emitters at lower redshifts could become contaminants and these could be 
obstacles in the application of this method.  An advantage of the 21cm line for 
intensity mapping is that due to its low frequency (1420 MHz),  
there are few strong lines at a lower redshift which could contaminate the 
observations\footnote{See e.g. \citet{Thompson01} (Table 1.1) for
a list of important radio spectral lines, the only one below HI 
frequency is the 327 MHz deuterium line, which is relatively 
weak due to the low deuterium abundance.}.

Nevertheless, the hydroxyl radical (OH) lines of 18 cm wavelength at slightly higher redshifts 
can potentially contaminate HI 21 cm observations at lower redshifts \footnote{Other lines 
at $\nu <10 \GHz$ listed in the above reference are the CH line at 3.335 GHz, 
the OH line at 4.766 GHz, the formaldehyde(${\rm H_2CO}$) line at 4.83GHz, 
the OH line at 6.035 GHz, the methanol(${\rm CH_3OH}$) line at 6.668 GHz, 
and the ${\rm ^3He}$ line at 8.665 GHz. These 
lines should be less significant than the OH 18cm line and we will not consider them in this
work.}. The $\lambda=$ 18 cm lines of OH correspond to four possible transitions, 
with frequencies at 1612, 1665, 1667, and 1720 MHz. Strong OH emission are 
produced by masers, originating typically in high 
density ($n(H_2) > 10^7 \cm^{-3})$ gas near an excitation 
source, though the exact environment for 
the masers to happen is still not clear \citep{Lo05}. 
The 1665MHz and 1667MHz are usually  much stronger than the other two lines, hence are named ``main lines''.
In masers, the 1667MHz line is the strongest, whose flux is typically about 
2 to 20 times greater than the 1665MHz line \citep{Randell95}.

In an intensity mapping observation, the 21 cm auto-correlation power spectrum at a redshift
$z$ is observed. However, the OH emission at
 $1+z'=(1+z)\frac{\lambda_{\rm HI}}{\lambda_{\rm OH}}$ would also 
give raise to brightness temperature fluctuations which can not be distinguished from 
the redshifted 21cm fluctuations in such observations.  Thus, for example, 
the HI 21-cm signal at $z_{\rm HI}=1,2$ and 3 would be 
contaminated by the OH 18cm emission 
at $z_{\rm OH}=1.35,2.52,$ and 3.70 respectively. 
Since the OH fluctuations at redshift $z'$ are 
uncorrelated with the 21 cm fluctuations at redshift
$z$, the two power spectra would simply add. 
Although the OH line emission is produced with a different 
mechanism and depends on the star formation activity, on large scales,
we still expect the OH intensity fluctuations to trace the total matter densities. Its power spectrum 
should be proportional to the matter power spectrum at $z'$, with a different 
bias factor. If not properly accounted for, this  may introduce a distortion to the total intensity power spectrum
extracted from  the 21 cm observations resulting in a shift to the BAO peaks. Given the low-redshift 21 cm BAO experiments are now being developed (e.g. the proposal to conduct wider area surveys with GBT by building a multi-beam receiver\footnote{https://science.nrao.edu/facilities/gbt/index}, the Tianlai project in China, and the CHIME project in Canada\footnote{http://www.phas.ubc.ca/chime/}) it is important to estimate the magnitude of the potential contamination.

For this purpose, we make use of simulations to predict both the 21 cm and OH intensity power spectrum from $z=1$ to 4.
We find that the contamination is generally small and below 1\% of the rms fluctuations at $z=1$ to 3 
with a large uncertainty related to the overall predictions on the OH signal.
The paper is organized as follows: In the next section we present the method of the intensity calculation and in Section~3 we present our results
and discuss the contamination. We will assume WMAP 7-year flat $\Lambda$CDM cosmological
model \citep{WMAP7}.

\section{Calculation}

In order to estimate the 21 cm emission at low redshifts, 
we make use of the semi-analytic simulations
by \cite{Obreschkow09}, which are available as part of the SKA Simulated Skies
\footnote{http://s-cubed.physics.ox.ac.uk}, and based on the galaxy catalog derived from 
the Millennium simulation \citep{De Lucia07,Springel05}. These are the same simulations we have used in 
\citet{Gong11}. As the HI mass of each galaxy is assigned using the galaxy properties provided by the semi-analytical modeling of galaxy formation, we will
use those neutral hydrogen masses to calculate the 21 cm line intensities.

The calculation of the 21 cm power spectrum is similar to what we have done
in \citet{Gong11}. 
The 21 cm temperature from galaxies, assuming the signal is seen in emission is \citep{Santos08}
\be
\bar{T}^{\rm G}_{\rm b} = c(z)\frac{\rho_{\rm HI}}{X_{\rm H}\rho_{\rm b}}\, (\rm mK),
\ee
where $X_{\rm H}=0.74$ is the mean hydrogen mass fraction in the Universe, 
$\rho_{\rm b}=\Omega_{\rm b}\rho_{\rm c}$ is the baryon density and $\rho_{\rm c}$
is the critical density. The $c(z)$ takes the form as
\be
c(z) \approx 23\bigg(\frac{0.7}{h}\bigg)\bigg(\frac{\Omega_{\rm b}h^2}{0.02}\bigg)
\bigg(\frac{0.15}{\Omega_{\rm m}h^2}\frac{1+z}{10}\bigg)^{1/2} ({\rm mK}).
\ee
The parameter $c(z)\bar{x}_{\rm H}$ gives the mean brightness 
temperature of the 21-cm emission, where $\bar{x}_{\rm H}$ is
the mean neutral fraction. 
We assume that the neutral hydrogen are mostly 
contained within the galaxies after reionization, the mass density $\rho_{\rm HI}$ is 
then given by
\be
\rho_{\rm HI} = \int_{M_{\rm min}}^{M_{\rm max}} dM \frac{dn}{dM}M_{\rm HI}(M),
\ee
where $dn/dM$ is the mass function,
we take $M_{\rm min}=10^8\ \rm M_{\odot}/h$ to be the minimum mass for 
a halo to retain neutral hydrogen \citep{Loeb01}, and $M_{\rm max}=10^{13}\ \rm M_{\odot}/h$
is the maximum mass for which the gas have sufficient time to cool and form 
galaxies (the result is insensitive to this number). 

In above the $M_{\rm HI}$ is the neutral hydrogen mass in a halo with mass $M$.
The HI mass is correlated with the halo mass, though with some scatter.
Inspired by the shape of the distribution seen in the
semi-analytic simulation generated from \cite{Obreschkow09}, we fit a relation of the form
\begin{equation}
M_{\rm HI}(M)=A\times \left( 1+\frac{M}{c_1}\right)^b \left( 1+\frac{M}{c_2}\right)^d \, .
\label{eq:MHI-M}
\end{equation}
The best-fit values of the parameters $A$, $c_1$, $c_2$, $b$ and $d$
are given in Table \ref{tab:MHI_M_fit}. 

\begin{figure}[t]
\begin{center}
\includegraphics[scale=0.413]{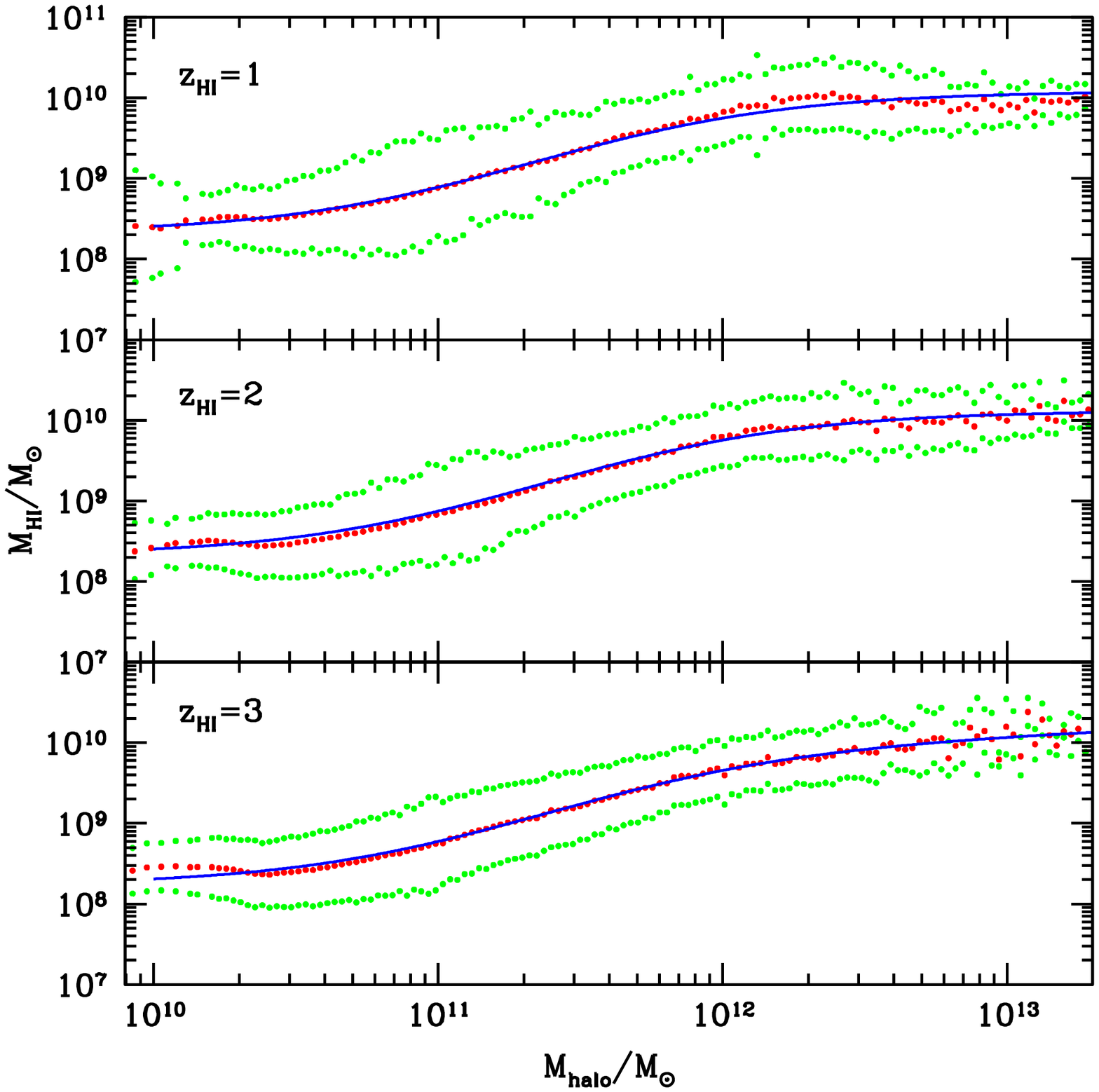}
\includegraphics[scale = 0.43]{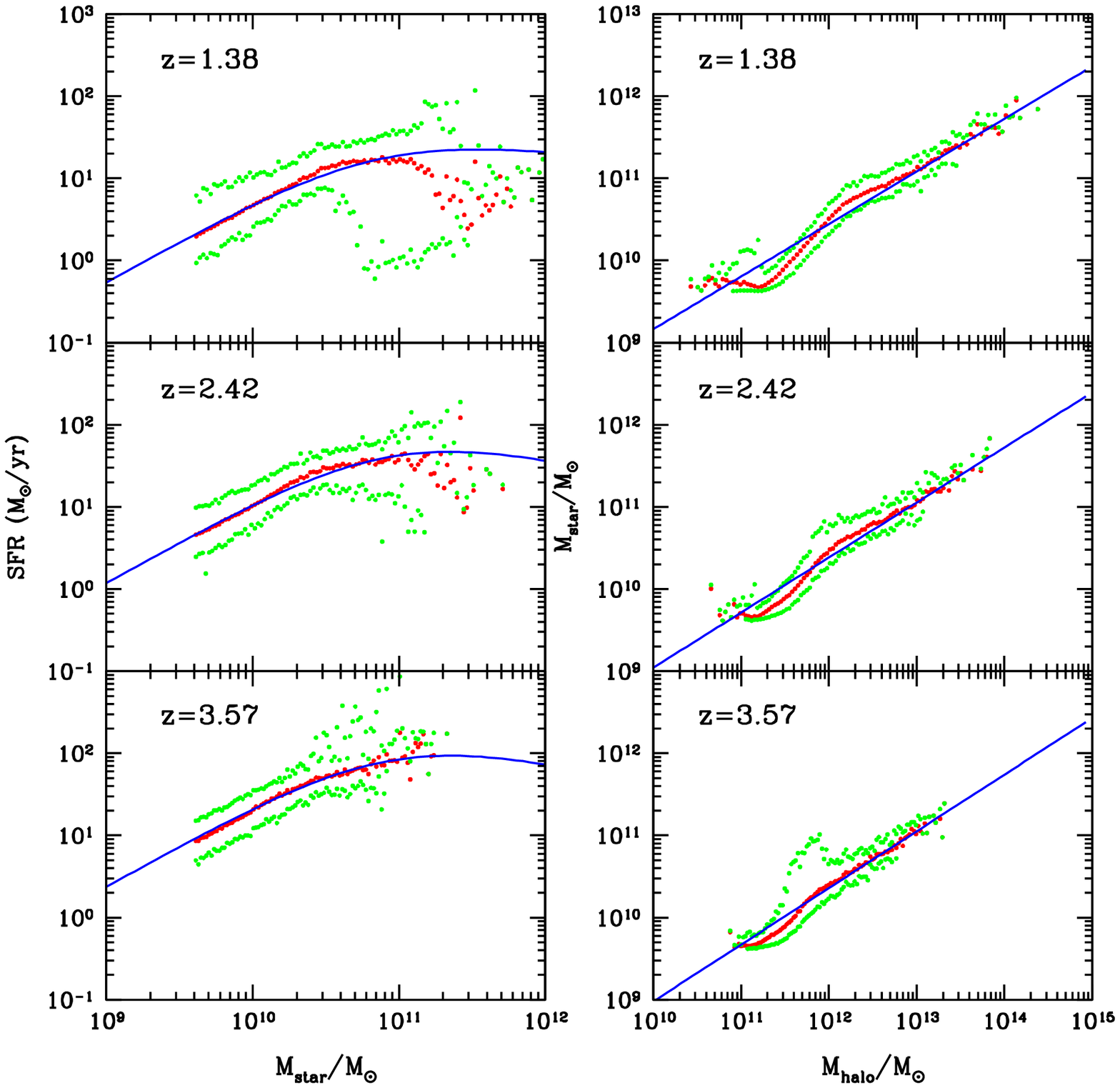} 
\caption{\label{fig:MHI_SFR} Upper panel: the $M_{\rm HI}$-$M$
relation from the simulation at $z=1$, $z=2$ and $z=3$. 
Lower panel: the SFR vs. the stellar mass $M_\star$ and the stellar 
mass vs. the halo mass at different redshifts as derived 
from the galaxy catalog in \cite{De Lucia07}. The red dots are the mean
value and the green dots show the 1$\sigma$
error from the simulation. The best fits are shown in blue solid line.}
\end{center}
\end{figure}

Due to the mass resolution limit of the 
Millennium simulation, for $M<10^{10}\ M_{\odot}$, we cannot use the same fitting
formula. Instead we assume $M_{\rm HI}=X_{\rm HI}^{\rm gal}
(\Omega_b/\Omega_m)M$ to estimate the neutral hydrogen mass in halos, 
where $X_{\rm HI}^{\rm gal}$ is the neutral hydrogen mass fraction in the galaxy. 
We set $X_{\rm HI}^{\rm gal}=0.15$
which is estimated at $M=10^{10}\ M_{\odot}$ in the simulation,
and assume it does not change when $M<10^{10}\ M_{\odot}$.
The simulation result and the best fitting curves at $z=1$, $z=2$ and $z=3$ are shown 
in the upper panel of Fig.~\ref{fig:MHI_SFR}, which are consistent with the other results 
(e.g. \cite{Marin10,Duffy11}). 

We find the HI energy density parameter $\Omega_{\rm HI}=\rho_{\rm HI}/\rho_c$ 
is about $10^{-3}$ and insensitive to the redshift (for $z\lesssim 3$) in our
calculation, which is consistent with the observational results
(e.g. \cite{Rao06,Lah07,Noterdaeme09}). Finally, we find
the 21-cm mean brightness temperature are 481, 573 and 544 $\mu$K at z=1, 2 and
3 respectively. These values are also consistent with an observation at $z=0.94$ \citep{Chang10} and
previous predictions in the literature \citep{Chang08}.

Assuming that the 21 cm flux from galaxies is proportional to the neutral hydrogen 
mass $M_{\rm HI}$, then the 21 cm signal will follow the underlying dark matter distribution with a bias
\be
b_{\rm HI}(z) = \frac{\int_{M_{\rm min}}^{M_{\rm max}} dM \frac{dn}{dM} M_{\rm HI} b(z,M)}{\rho_{\rm HI}}.
\label{eq:bias}
\ee
where $b(z,M)$ is the halo bias \citep{Sheth99}. The 21-cm temperature from galaxies is then
$T^{\rm G}_{\rm b}({\bf x}) = \bar{T}^{\rm G}_{\rm b}[1+b_{\rm HI}\delta({\bf x})]$,
and the clustering power spectrum is given by 
$P^{\rm G}_{\rm HI}(z,k) = \left(\bar{T}^{\rm G}_{\rm b}\right)^2 b^2_{\rm HI}P(k,z).$
We use the $\tt Halofit$ code \citep{Smith03}
to calculate the non-linear matter power spectrum $P(k,z)$. 
Additionally, there is a shot noise contribution to the power spectrum due to the 
discreteness of galaxies, 
\be
P^{\rm shot}_{\rm HI}(z) = \int_{M_{\rm min}}^{M_{\rm max}} dM \frac{dn}{dM}
\left[c(z)\frac{M_{\rm HI}}{X_{\rm H}\rho_{\rm b}} \right]^2.
\ee
\begin{table}[t]
\centering
\caption{The best-fit values of the parameters in the $M_{\rm HI}$-$M$
relation.}     
\begin{tabular}{c | c c c}        
\hline\hline                 
  & $z=1$ & $z=2$ & $z=3$ \\
\hline
$A$ & $2.1\times 10^{8}$ & $2.1 \times 10^{8}$ & $1.7 \times 10^{8}$\\
$c_1$ & $1.0\times 10^{11}$ & $1.0\times 10^{11}$ &  $1.0\times 10^{11}$ \\
$c_2$ & $4.55\times 10^{11}$ & $5.60\times 10^{11}$ &  $5.0\times 10^{11}$ \\
$b$  & $2.65$ & $2.4$ &  $2.4$ \\
$d$ & $-2.64$ & $-2.40$ & $-2.25$ \\
\hline \hline                               
\end{tabular}
\label{tab:MHI_M_fit}     
\end{table}

While the OH molecule is not as abundant as neutral hydrogen, the OH maser is
brighter than the 21 cm line intensity for the same number density of baryons. It is believed that OH masers are 
associated with star formation activity. OH megamasers (OHMs)  
are $10^6$ times brighter than typical OH maser sources within the Milky Way and are
found in the luminous infrared galaxies 
(LIRGs with $L_{\rm IR}>10^{11}\ L_{\odot}$) and ultra-luminous infrared galaxies 
(ULIRGs with $L_{\rm IR}>10^{12}\ L_{\odot}$) \citep{Darling02}. The luminosity 
of the isotropic OHM luminosity $L_{\rm OH}$ is correlated with the IR luminosity 
$L_{\rm IR}$.

Previous studies found several relations that take the form as
$L_{\rm OH}\propto L_{\rm IR}^{\beta}$, where the power index $\beta$ is between 1 and 2.
For example, \cite{Baan89} found a relation $L_{\rm OH}\propto L_{\rm IR}^2$ 
using a sample of 18 OHM galaxies. However, a
flatter relation of $L_{\rm OH}\propto L_{\rm IR}^{1.38}$ was obtained by \cite{Kandalian96} 
using a sample of 49 OHM galaxies, after correcting for the Malmquist bias.
A nearly linear relation between $L_{\rm OH}$ and $L_{\rm IR}$ was derived 
from the Arecibo Observatory OH megamaser survey \citep{Darling02}:
\be \label{eq:L_OH_L_IR}
{\rm log}L_{\rm OH}=(1.2\pm0.1){\rm log}L_{\rm IR}-(11.7\pm1.2).
\ee
This relation has also been corrected for the Malmquist bias, and about one hundred OHM galaxies
are used in this calibration. We will use this relation in our model.

The $L_{\rm IR}$ is tightly correlated with the star formation rate (SFR)
and we adopt a relation of the form \citep{Magnelli11,Tekola11}
\be \label{eq:L_IR_SFR}
L_{\rm IR}\ [L_{\odot}]=5.8\times 10^9\ {\rm SFR}\ [\rm M_{\odot}yr^{-1}].
\ee
This relation is consistent with other works (e.g. \cite{Evans06}), and
have about $30\%\sim40\%$ uncertainty \citep{Kennicutt98,Aretxaga07}.

The SFR is on a statistical sense linearly correlated with the halo 
mass \citep{Loeb05,Shimasaku08}, with a nearly Gaussian distribution whose central value
and variance changes as a function of redshift \citep{Conroy09}.
For the purpose of statistical calculation of OH emissivity, it is sufficient to 
relate the star formation rate to the halo mass.
We use the galaxy catalog in \cite{De Lucia07} to derive the SFR and stellar mass relation SFR-$M_{\star}$ and 
the star mass and halo mass relation $M_{\star}$-$M$ as shown in the lower panel of Fig.~\ref{fig:MHI_SFR}. 
The simulation has outputs at $z=1.38$, 
$2.42$ and $3.57$, which are fairly close to the redshifts $z_{\rm OH}=1.35$, 
$z_{\rm OH}=2.52$ and $z_{\rm OH}=3.70$, which could contaminate the 21cm 
at $z_{\rm HI}=1$, $z_{\rm HI}=2$ and $z_{\rm HI}=3$ respectively.

We fit the SFR-$M_\star$ and $M_\star$-$M$ relations using the form as 
${\rm SFR}=A\times M_\star(1+M_\star/c)^d$ and $M_\star=B\times M^e$ respectively,
and the best fit values for the parameters are listed in Table \ref{tab:SFR_M_fit}.
For $M<10^{12} M_{\odot}$, analysis from the simulation indicates 
SFR$/M \sim 10^{-11}$-$10^{-10}\ \rm yr^{-1}$ at $1<z<4$, which matches well with previous
results\citep{Loeb05,Shimasaku08,Conroy09} in their applicable redshift ranges.

\begin{table}[htbp]
\centering                    
\caption{The best-fit values of the parameters in the SFR-$M_{\star}$
and $M_{\star}$-$M$ relation.}
\begin{tabular}{c | c c c}        
\hline\hline                 
  & $z=1.38$ & $z=2.42$ & $z=3.57$ \\
\hline
$A$ & $5.5\times 10^{-10}$ & $1.2\times 10^{-9}$ & $2.4\times 10^{-9}$ \\
$c$ & $7\times10^{10}$ & $9\times10^{10}$ & $9\times10^{10}$ \\
$d$ & $-1.2$ & $-1.4$ & $-1.4$ \\
\hline
$B$ & $5.8\times 10^2$ & $2.2\times10^2$ & $1.2\times10^2$\\
$e$ & $0.64$ & $0.67$ & $0.69$ \\
\hline \hline                   
\end{tabular}
\label{tab:SFR_M_fit}     
\end{table}

\begin{figure}[t]
\begin{center}
\includegraphics[scale = 0.43]{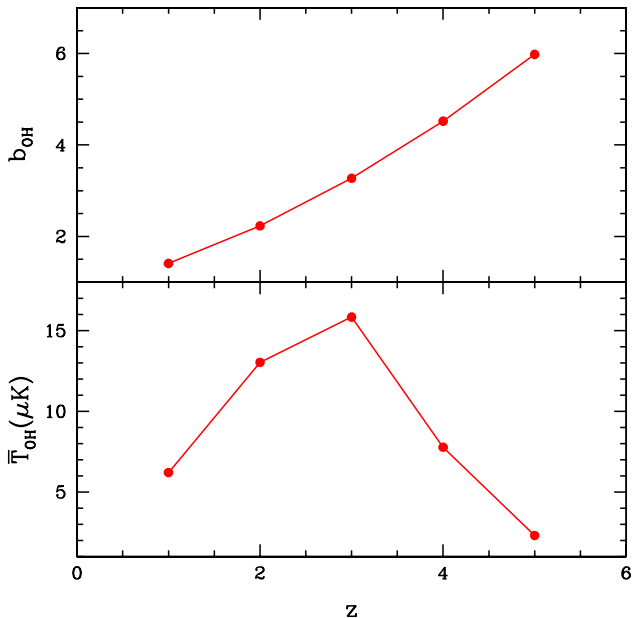} 
\caption{\label{fig:b_T_OH} The OHM clustering bias and the
mean Rayleigh-Jeans temperature at z=1, 2, 3, 4 and 5.
We find the OHM signal declines quickly when $z>3$.
}
\end{center}
\end{figure}

We can now calculate the mean intensity of the OH emission:
\be
\bar{I}_{\rm OH}(z) = f_{\rm OH}\int_{M_{\rm min}}^{M_{\rm max}}dM\frac{dn}{dM}f_{\rm IR}(M)\frac{L_{\rm OH}(M,z)}{4\pi D_L^2}y(z)D_A^2,
\ee
where the $dn/dM$ is the halo mass function \citep{Sheth99}, $D_L$ and $D_A$ are the luminosity
distance and comoving angular diameter distance respectively, and 
$y(z)=d\chi/d\nu=\lambda_{\rm OH}(1+z)^2/H(z)$, 
where $\chi$ is the comoving distance, $\nu$ is the observed frequency, 
$\lambda_{\rm OH}=18\ \rm cm$ is the rest-frame OH wavelength.
The $f_{\rm OH}$ is the fraction of the LIRGs together with ULIRGs
that host the OHMs. We set $f_{\rm OH}=0.2$ \footnote{This value also can 
be as low as 0.05, see \cite{Klockner04}.} which is estimated using
the sample in \cite{Darling02}.  This takes account of the fact that the OHMs are caused by 
the far-IR pumping ($53\ \rm \mu m$) in the warm dust ($T>45\ \rm K$) which is 
produced and supported by the star formation in LIRGs or ULIRGs \citep{Lockett08}.
Note that the duty cycle does not appear in our formula, since it is already
included in our SFR-$M$ relation.

Here we assume that the megamasers would dominate the contribution, so 
we can take $M_{\rm min}=10^{11}\ \rm M_{\odot}/h$, as the OHMs
are hosted in galaxies with molecular gas $M_{\rm H_2}\gtrsim 4\times 
10^9\ \rm M_{\odot}$ \citep{Burdyuzha90,Lagos11,Duffy11}. The $f_{\rm IR}(M)$
is the fraction of the LIRGs and ULIRGs for galaxies hosted by the halos 
with $M>10^{11}\ \rm M_{\odot}/h$, which is estimated from the catalog in 
\cite{De Lucia07}. We find $f_{\rm IR}\sim 1$ when $M>1.2\times10^{12}\ \rm M_{\odot}$,
and quickly decreases for lower halo masses.
After getting the OHM intensity, we can
convert it into a mean Rayleigh-Jeans temperature $\bar{T}_{\rm OH}$.

The OHM clustering bias can be calculated in the same way as the HI bias
(c.f. Eq.~\ref{eq:bias}), except for the
weight by $L_{\rm OH}$ instead of $M_{\rm HI}$. The OHM bias $b_{\rm OH}$ and
$\bar{T}_{\rm OH}$ at z=1, 2, 3, 4 and 5 are shown in Fig.~\ref{fig:b_T_OH}.
We find that the OHM signal declines quickly when $z>3$.

The OHM power spectrum is given by 
$P_{\rm OH}(k,z) = \bar{I}_{\rm OH}^2b_{\rm OH}^2P(k,z)$,
and similar to the HI case, the shot-noise power spectrum is given by
\ba
P_{\rm OH}^{\rm shot}(z) &=& f_{\rm OH}\int_{M_{\rm min}}^{M_{\rm max}}dM\frac{dn}{dM} \nonumber \\ 
                         &\times& f_{\rm IR}(M)\left[\frac{L_{\rm OH}(M,z)}{4\pi D_L^2} y(z)D_A^2\right]^2.
\ea

\begin{figure}[t]
\begin{center}
\includegraphics[scale = 0.43]{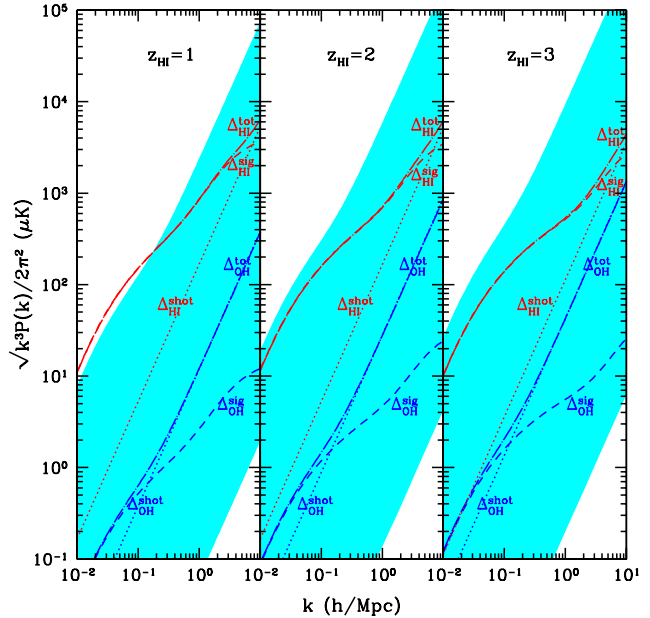} 
\caption{\label{fig:rms_OH_21} The rms intensity fluctuations using the power spectrum of 
the OHM and 21-cm emission. The blue dashed, dotted 
and dash-dotted lines are the OHM signal, shot-noise 
and total power spectrum. The uncertainty of the OHM
emission is also shown in cyan region,
which is estimated using the errors in Eq.\ref{eq:L_OH_L_IR} only.
The red lines are 21-cm power spectrum.
Because the shot-noise power spectrum is too
small compared to the signal power spectrum
for 21-cm, the total and the signal power spectrum
are almost overlapped together.
}
\end{center}
\end{figure}

\section{Results and Discussion}

In Fig.~\ref{fig:rms_OH_21}, we plot the rms fluctuations associated with the power spectrum of the OH and
21-cm emission at different redshifts for comparison. 
The 21-cm signal is plotted in red color, while the OH in blue.
In the case of 21 cm power spectrum, the shot noise power is relativity insignificant, 
as the number of HI galaxies is large.

In the case of the OH power spectrum, the signal power spectrum evolves slowly in this 
redshift range,  because the SFR is higher at high redshifts, so it counteracts the decrease in the 
matter power spectrum at high redshifts. The contribution of the shot noise 
is however very significant, especially at small scales (larger $k$), as the 
emission is mostly from the rare LIRG and ULIRG populations.  

Comparing the 21cm power to the OH power, we find that on the scale of the first 
peak of the BAO (about $k=0.075 h/\Mpc$), the OH rms fluctuations are about $0.03\%$, $0.07\%$
and $0.11\%$ of the 21 cm rms fluctuations at $z_{\rm HI}=1$, 2 and 3 
respectively. At higher redshifts, while we do not show it here, we found that the 21cm 
signal become stronger as we approach the epoch of reionization, while the OH 
power become smaller and insignificant compared with the HI signal. 

We note that this result depends on the modeling of the HI and 
OH emission, which still has a lot of 
uncertainty. The exact conditions for the occurrence of OH megamasers 
are not completely understood \citep{Lo05}, and the 
actual OH emission from a source of a certain star formation rate
may be quite different from our model prediction. Moreover,
the OH emissivity may not even be strongly correlated with the halo mass, 
though on very large scales we still expect that  the OH intensity
power spectrum to be proportional to the underlying matter power spectrum. The 
21 cm power depends on the HI content of the low mass halos, which is also largely uncertain.

Considering variations to our model predictions, we do find that in an extreme case, 
as shown the upper limit of the cyan region in Fig.~\ref{fig:rms_OH_21},
the OH could even supersede the 21cm power spectrum, and become the 
major contribution to the observed temperature fluctuations. 
Note that we just consider the errors in the $L_{\rm OH}$-$L_{\rm IR}$ relation 
(Eq.~\ref{eq:L_OH_L_IR}) to get
the uncertainty (cyan region), which can be greater if including the error
in $L_{\rm IR}$-SFR relation (Eq.~\ref{eq:L_IR_SFR}). Also, we note that
the two relations above are calibrated at low redshift and tightly related to
the redshift-dependent properties of galaxies, such as the galaxy metallicity.
So the OH intensity may also increase if considering the redshift evolution effect.
Of course, in that case, it would be more advantageous to  
use the OH emission as the tracer instead, though
at present this does not seem to be very likely. 

As a lot of the OH power comes from shot noise, it may be possible to find a way to remove
some of its contribution. For example, we may consider to conduct a targeted  maser survey on the 
ULIRGs and LIRGs with sensitive telescopes which have small fields of view, and subtract their
contributions to the temperature fluctuation. This could significantly 
reduce the noise power due to these sources. 
 
One way to identify and estimate the amount of possible contamination is to 
cross correlate the temperature fluctuation at the redshift pair $(z_1, z_2)$.
We expect the temperature fluctuations should be uncorrelated, while contamination 
by OH would give raise a correlation given by 
$\langle \delta T(k, z_1) \delta T(k,z_2) \rangle =b_{\rm HI}(z_2) b_{\rm OH}(z_2) P(k,z_2)$.

Finally, it may also be possible to make use of the multiple lines 
of OH (including the four lines at 18cm and the lines with shorter wavelengths) 
to check for the contamination. If observations for individual OHM sources show 
that most of them have similar line ratios, then one may construct a template of OHM 
spectrum, and apply it as a matched filter on the observed spectrum to check for possible
OH contamination. However, this would not be possible if observations show that 
the line ratios vary a lot.

We thank Prof. Xingwu Zheng for helpful discussion. This work was supported by 
NSF CAREER AST-0645427 at UCI, by the 973 program 
No. 2007CB815401, the NSFC grant No.11073024, and the John Templeton Foundation at
NAOC; and MGS and MBS acknowledges support from FCT Portugal
under grant PTDC/FIS/100170/2008.



\end{document}